\documentclass[aps,preprintnumbers,prl,twocolumn,nofootinbib]{revtex4}
\usepackage{epsfig,latexsym,cancel,amssymb,amsmath,mathrsfs}
\usepackage{graphicx}
\usepackage{epstopdf}
\usepackage{mciteplus}
\usepackage{latexsym}
\usepackage{amsthm}
\usepackage{amsmath}
\usepackage{amssymb}
\usepackage{hepunits}
\usepackage{hyperref}
\usepackage{bbm}
\usepackage{bm}
\usepackage{xfrac}
\usepackage{color}
\usepackage{colordvi}
\usepackage{comment}
\usepackage{dcolumn}
\usepackage{times,latexsym,graphicx,wrapfig}
\usepackage{epsfig,lineno,bm}
\usepackage{subfigure}
\usepackage{slashed}

\usepackage{calrsfs}
\DeclareMathAlphabet{\pazocal}{OMS}{zplm}{m}{n}

\newcommand{\be}{\begin{eqnarray}}
\newcommand{\ee}{\end{eqnarray}}
\newcommand{\bea}{\begin{eqnarray}}
\newcommand{\eea}{\end{eqnarray}}

\newcommand{\bef}{\begin{figure}[htbp]\begin{center}}
\newcommand{\eef}{\end{center}\end{figure}}

\newcommand\FNAL{Fermi National Accelerator Laboratory, Batavia, IL, 60510, USA}

\def\lsim{\mathrel{\rlap{\lower4pt\hbox{\hskip1pt$\sim$}}
\raise1pt\hbox{$<$}}}
\def\gsim{\mathrel{\rlap{\lower4pt\hbox{\hskip1pt$\sim$}}
\raise1pt\hbox{$>$}}}

\begin{document}

\preprint{FERMILAB-PUB-16-110-PPD}
%\hspace{4.5cm}~~{\scriptsize FERMILAB-PUB-16-110-PPD}  

%\title{ Is the Baryon Asymmetry a Pre-existing Condition?}
\title{ Can the  Baryon Asymmetry Arise From Initial Conditions?}
\author{ Gordan~Krnjaic}   
 \affiliation{\FNAL}

\begin{abstract}
In this letter, we quantify the challenge of explaining the baryon asymmetry using initial conditions in a universe
that undergoes inflation. Contrary to lore, we find that such an explanation is possible if net $B-L$ number is stored in a light bosonic field with  hyper-Planckian initial displacement and a delicately chosen field velocity prior to inflation.  However, such a construction may require extremely tuned  coupling constants to ensure that this asymmetry is viably communicated to the Standard Model after reheating; the large field displacement required to overcome inflationary dilution must not induce masses for Standard Model particles or generate dangerous washout processes. While these features are  inelegant, this counterexample nonetheless 
 shows that there is no theorem against such an explanation. We also comment  on potential observables in the double $\beta$-decay spectrum and on
 model variations that may allow for more natural realizations.
\end{abstract}

\maketitle

%%%%%%%%%%%%%%%%%%%%%%%%%%%%%%%%%%%%%%%%%%%
%							Introduction
%%%%%%%%%%%%%%%%%%%%%%%%%%%%%%%%%%%%%%%%%%%

\section{Introduction}
The observed flatness, isotropy, and homogeneity of the universe on large scales strongly suggest that the universe  
underwent a period of inflation at early times (for a review see \cite{Dine:2003ax}). An important consequence of this epoch is that pre-inflationary
relics are exponentially diluted away, thereby eliminating monopoles, cosmic strings, and other topological
defects from the observed universe. Such a dilution also depletes
any primordial $B-L$ number,  so it is generally accepted that a dynamical mechanism satisfying the Sakharov conditions \cite{Sakharov:1967dj} is required
to generate  the matter asymmetry after inflation. 

To see the difficulty, consider the most favorable initial condition for preserving an asymmetry stored in Standard Model (SM) particles: a universe populated with a degenerate gas of massless, asymmetric neutrinos ($n_\nu > n_{\bar \nu} = 0$) at zero temperature.  Denoting pre-inflationary
quantities with an $i$ subscript, the 
energy density of this population is
 \be \label{eq:fermi}
 \rho_{\nu,i} =   \frac{1}{4\pi^2}  (3 \pi^2  n_{\nu,i} )^{4/3}~,~
 \ee
which is completely determined by the number density  $n_{\nu,i}$, so after 
  ${N}$ inflationary e-folds, the diluted asymmetry is
  \be\label{eq:Ynu}
  Y_\nu \equiv \frac{n_{\nu}}{s} = \frac{45        (4 \pi^2 \rho_{\nu,i})^{3/4}   }{6 \pi^4  g_* T_R^3 } \, e^{-3N}~,~
  \ee
where  $s$ is the entropy density,  $T_R$ is the reheating temperature,
 and $g_*$ is the number of relativistic degrees of freedom. 
Even with a Planckian energy density and the lowest viable reheating temperature the asymmetry is at most 
\be
Y^{\rm max}_\nu  \sim  4\times 10^{-15} \left( \frac{\rm 10\, MeV}{ T_R} \right)^3 \left( \frac{\rho_{\nu,i}}{ m_{Pl}^4 } \right)^3 ,
\ee
so even if $Y^{\rm max}_\nu$ is all converted into a comparable baryon asymmetry, it falls many orders of magnitude below the observed value
$Y^{\rm obs}_{B} \equiv (n_B-n_{\bar B}) /s = (8.6 \pm 0.1)\times 10^{-11}$ \cite{Ade:2013zuv}.
It is also clear that, for an initial asymmetry of order  $\sim Y^{\rm max}_\nu$, the corresponding energy density $\rho_{\nu,i}$ easily dominates the pre-inflationary universe, thereby preventing inflation from starting in the first place.

However, this observation assumes that the number density is stored in fermions for which the energy and number densities are related by Fermi
 statistics as in Eq.~(\ref{eq:fermi}). If, instead, the asymmetry is stored in a bosonic field, these quantities may be decoupled from each other and it is possible to engineer  a relatively low energy configuration with a large number density.
 
  In this letter, we study the conditions under which a large $B-L$ asymmetry can exist
 prior to inflation, survive $\sim 60$ e-folds of dilution, and viably yield the observed baryon asymmetry after reheating.  
 We find that satisfying these conditions is possible, but requires super-Planckian field values and significant tuning to stabilize the electroweak sector and prevent the asymmetry from being washed out.  Nonetheless, the simple realization presented here serves as a proof of  principle to demonstrate that a sizable asymmetry can, indeed, survive inflation and be transferred to SM particles.  Although \cite{Dolgov:1994zq} initially suggested the possibility of storing an asymmetry in a light boson and transferring it to the SM, to our knowledge this work presents 
the  first explicit model to realize such a mechanism in a universe with inflation. 
  
%%%%%%%%%%%%%%%%%%%%%%%%%%%%%%%%%%%%%%%%%%%
%							Initial Setup
%%%%%%%%%%%%%%%%%%%%%%%%%%%%%%%%%%%%%%%%%%%

\section{Model Description and Initial Setup}
Inspired by the Affleck-Dine mechanism \cite{Affleck:1984fy}, we consider a complex scalar that carries a global conserved $\phi$ number, which we will later identify with $B-L$.\footnote{Although quantum gravity may spoil this global symmetry, we assume that the relevant dynamics occur at scales for which these effects can be neglected.} The Lagrangian for this setup is
\be
{\mathcal L} =  |\partial_\mu \phi|^2 - m_\phi^2 |\phi|^2 - V(|\phi|) ~,~
\ee
which preserves $\phi$ number in its potential. Defining $\phi = \frac{1}{\sqrt{2}}r e^{i\theta}$, this can be written in polar coordinates as 
\be
{\mathcal L} =  \frac{1}{2}     \left(  \dot r^2 + r^2 \dot \theta^2 - m_\phi^2 r^2  \right)  - V(r)~.~~
\ee
The  $\phi$ number density is  
\be\label{eq:numberdensity}
n_{ \phi} =  i(\phi^* \partial_t \phi - \phi \, \partial_t\phi^*) =  r^2 \dot \theta ~,~~
\ee
so  the angular equation of motion can be written as 
\be\label{eq:phi-eom}
a^{-3}\,\partial_t (a^3 n_{\phi} ) =  - \partial_\theta V =  0~,~~
\ee
where $a$ is the FRW scale factor.
Like in Affleck-Dine models,  we assume that $\phi$ is initially displaced far from the origin in field space.
 Unlike in Affleck-Dine, this displacement is established {\it before} inflation and we 
allow no explicit $\phi$ number violation in the Lagrangian ({\it i.e.} the potential is independent of $\theta$)  so Eq.~(\ref{eq:phi-eom}) represents the conservation of comoving $\phi$ number.

 To generate a sizable asymmetry from initial conditions, we require that  
\begin{itemize}
\item  The asymmetry in  $\phi$-number before inflation must be exponentially large to survive 60 e-folds of inflation, which implies $n_{\phi,i} = r_i^2 \dot \theta_i \gg m_{Pl}^3$.
\item  This large number density must not dominate the the energy budget of the pre-inflationary universe, $\rho_{\phi,i} <  \rho_{\rm inf}$, so that 
the inflaton's potential can successfully drive inflation in the presence of such a large number density.
\item $\phi$ number must consistently be identified with $B-L$ and that the corresponding asymmetry must be transferred to the SM through $B-L$ preserving
interactions. 
 \end{itemize}
  
\noindent Naively, these are difficult tasks since values of $n_{\phi,i}$ that survive 60 e-folds require very large field values, which generically favor a comparably large energy density 
\be\label{eq:generic-energy-density}
\rho_{\phi,i} &=&  \frac{1}{2}     \left(  \dot r_i^2 + r_i^2 \dot \theta^2_i + m_\phi^2 r_i^2  \right)  + V(r_i) ~.~
\ee
 However, consider initial conditions in which 
  \be\label{eq:IC}
  r_i  \gg m_{Pl}  ~~~,~~~    \dot r_i \ll \sqrt{\rho_{\rm inf}}        ~~~,~~~             m_\phi   \ll   \dot \theta_i \ll \frac{   \sqrt{\rho_{\rm inf }}          }{~r_i}  ~~,~~
  \ee
 where $\rho_{\rm inf}$ is the energy density during inflation. In this limit, $\phi$ is relativistic prior to inflation and the physical energy density is dominated by the angular momentum term, so we have
 \be\label{eq:rho-limit}
\rho_{\phi,i} &\simeq& \frac{r_i^2 \dot\theta_i^2}{2} =   \frac{  n_{ \phi,i} \, \dot \theta_i}{2}  \ll  \rho_{\rm inf} ~,~
\ee
 where the contributions from $V(r_i)$ and $m_\phi^2 r_i^2$ have been tuned away by the bare cosmological constant. Note that the number 
 density $n_{\phi,i} = r_i^2 \dot \theta_i$ can be very large while the 
 average energy per particle $\rho_{\phi,i}/n_{\phi,i} \sim \dot \theta_i $ can be small for suitable choices of $r_i$ and $\dot \theta_i$.
  Note also that $m_\phi$ must be negligible compared to all other scales in the problem because, for a non-relativistic ensemble, 
the energy density $(\rho_\phi \ge m_\phi n_\phi)$ cannot be decoupled from the number density.
 However, if $\phi$ is the Nambu-Goldstone boson (NGB) of a broken global symmetry, its mass is protected by a shift symmetry and  can naturally be light; we will return
to this possibility below.  
 
Assuming instantaneous reheating to temperature $T_R$ immediately following inflation, the post-inflationary $\phi$ asymmetry  is
\be
Y_{\phi} &=& \frac{n_\phi  }{ s} = \frac{ 45 \, e^{-3N}     \,   n_{\phi,i}     }{ 2 \pi^2   g_{*}T_R^3}\, ,~
\ee
 where $g_*$ is the number of relativistic degrees of freedom.
Requiring ${N} = 60$ e-folds during inflation, we find 
\be\label{eq:Yphi} 
Y_{\phi}\simeq    10^{-10}  \left(  \frac{53 \, {\rm TeV}}{T_R} \right)^3  
                             \left(  \frac{r_{i}}{10^{67} {\rm \, GeV}} \right)^2  
                              \biggl(  \frac{\dot \theta_{i}}{10^{-50} {\rm \, GeV}} \biggr),~~~~~~~
\ee
where we have taken $g_* = 100$. 
Although the initial field value $r_i$ and angular velocity $\dot\theta_i$ have taken on grotesquely hierarchical values, the asymmetry in $\phi$ is 
in the right ballpark and the energy density in Eq.~(\ref{eq:rho-limit}) remains sub-Planckian
\be \label{eq:rho}
\rho_{\phi,i}  \simeq  2 \times 10^{-43}\, m_{Pl}^4    
\left(  \frac{r_{i}}{10^{67} {\rm \, GeV}} \right)^2  
                              \biggl(  \frac{\dot \theta_{i}}{10^{-50} {\rm \, GeV}} \biggr)^2\!\!,~~~~~~
\ee
which, therefore, allows the inflaton potential to dominate the total energy density at early times.  

%%%%%%%%%%%%%%%%%%%%%%%%%%%%%%%%%%%%%%%%%%%
%					Field Evolution Post Inflation
%%%%%%%%%%%%%%%%%%%%%%%%%%%%%%%%%%%%%%%%%%%

\section{Field Evolution Post-Inflation}
To understand how the field evolves during and after inflation, we solve the classical equations of motion  
\be
\label{eq:radial}     a^{-3} \partial_t ( a^3 \dot r)  = m_\phi^2 r - r \dot \theta^2 - \partial_r V \simeq 0 ~~~~~ \\ 
\label{eq:angular}   \partial_t ( a^3 r^2 \dot \theta) =- \partial_\theta V =0~,~~~~~~
 \ee
 where, based on Eq.~(\ref{eq:IC}), the $m_\phi^2 r$ and $r \dot \theta^2$ terms are negligible in Eq.~(\ref{eq:radial}).  We also assume
 that $|\partial_r V|^{1/3} \ll H, \rho^{1/4}_{\rm inf}$ during inflation so that the radial dynamics are dominated by Hubble expansion; since $r_i$ is hyper-Planckian and $\rho_\phi \ll \rho_{\rm inf}$, the $\partial_r V$ term cannot significantly affect the radial displacement.  Finally, we demand that all interactions preserve $\phi$ number, so $\partial_\theta V=0$ in Eq.~(\ref{eq:angular}).  Aside from these
 requirements, the details of this potential are not important for our purposes.  Note that the initial values $r_i$ and $\dot \theta_i$ in Eq.~(\ref{eq:IC}) need not necessarily correspond to a minimum of the potential; they can be imposed by fiat when inflation begins.
     
 Introducing $c_r$ and $c_\theta$ as constants of integration with mass-dimension 1 and using the scale factor during
 inflation $a(t) \propto e^{Ht}$, the system in Eqs.~(\ref{eq:radial}) and (\ref{eq:angular}) becomes
 \be \label{eq:veldamp}
 \dot r \simeq c^2_r e^{-3Ht} ~~,~~
  \dot \theta = \frac{ c^3_\theta}{r_i^2} e^{-3Ht}~,~~
 \ee 
where $H\equiv \dot a/a$ is the Hubble rate and for $\dot \theta$ we have used the fact that $\dot r \sim e^{-3Ht}$, so the initial value $r_i$ is approximately preserved with Hubble expansion. In our regime of interest, Eq.~(\ref{eq:radial}) can be approximated as  $\partial_t ( a^3 \dot r  )  = 0$, which yields 
 \be\label{eq:sol}
%  r(t) &=& \frac{e^{-3Ht} }{3H} \left[     c^2_r \left( e^{3Ht}  -1   \right)  + 3H r_i e^{3Ht}   \right] \, \rightarrow  \, r_i ~~,~
     r(t) &=& r_i   +    \frac{c^2_r }{3H}      \left(1  - e^{-3Ht}   \right)    \, \rightarrow  \, r_i ~~,~
   \ee
where the limiting value holds after imposing initial conditions $\dot r(0)  = 0$ (equivalently $c_r = 0)$  and $r(0) = r_i$ from Eq.~(\ref{eq:IC}) .
There is a corresponding expression for $\theta(t)$ with the replacement $c^2_r \to c^3_\theta/r_i^2$ and $r_i \to \theta_i$, so the initial radial and
angular positions are quickly locked in when inflation begins. We have also verified numerically that including small corrections from $m^2_\phi r$, $r \dot\theta^2$, and $\partial_r V$ terms in Eq~(\ref{eq:radial}) does not affect the solution in Eq.~(\ref{eq:sol}) as long as our parametric assumptions 
hold.

For this setup, we can safely assume that the quantum fluctuations during inflation are negligible as $\langle |\delta \phi |^2\rangle \sim H^2 \ll r_i^2$ for our chosen initial configuration. Also since Eq.~(\ref{eq:rho}) implies $\rho_\phi \ll \rho_{\rm DM}$, isocurvature fluctuations are highly suppressed and do not 
affect CMB observables; below we will also require that $\phi$ thermalize 
with the SM at later times, so  such fluctuations would be erased. 

\section{Transferring the Asymmetry}
To communicate the conserved $\phi$ number to SM fields, we consistently assign $\phi$ a nonzero $B-L$ number $Q_\phi$ and introduce the $B-L$ preserving interaction
\be\label{eq:transfer-lepton}
{\cal L} \supset   \frac{1}{\Lambda^{D-2}}(\partial_\mu \phi)\,  \hat{\pazocal O}^\mu_{B-L} + {h.c.} ~,~
\ee
where $\hat{\pazocal O}_{B-L}$ is some SM operator with mass dimension $D$ and nonzero $B-L$ number such that $(\partial_\mu \phi) \,\hat {\pazocal O}^\mu_{B-L}$ is a $B-L$ singlet.  Note that non-derivative operators of the form $\phi \, \hat{\pazocal O}_{SM}$ communicate $B-L$ breaking to the SM through $r_i$ insertions, which must be sequestered from the SM to avoid washing out the primordial asymmetry and maintaining sub-Planckian SM energy densities. Achieving this separation may require tuning since there is no shift symmetry to forbid such interactions (see below for a discussion of this issue).    
After reheating populates the SM, the interaction in Eq.\,(\ref{eq:transfer}) thermalizes $\phi$ with SM particles, thereby transferring the net $B-L$ number stored in $\phi$ to the rest of the thermal bath. 
 
 One possible realization involves interpreting $\phi$ number as lepton number so that its asymmetry can be 
 transferred via 
\be\label{eq:transfer}
{\cal L}\! \supset \!  \frac{1}{\Lambda}(\partial_\mu \phi) {N^c}^\dagger \bar \sigma^\mu N \! +  M N N^c \! +    y_\nu H L N  \!+\!   h.c. ,~~~~~~~~              
\ee
where flavor indices have been suppressed and the SM has been extended to include gauge singlet $N$, its Dirac partner $N^c$, and the scalar $\phi$ whose $B-L$ numbers are $+1$, $-1$ and $-2$, respectively so that all terms in Eq.~(\ref{eq:transfer}) preserve $B-L$. 
%flipped signs to make it B-L number instead of just L number
Note that the current-like operator $ {N^c}^\dagger \bar \sigma^\mu N$ is not the usual RH neutrino vector current and that $MN N^c$ is a {\it Dirac} mass, where we assume $\Lambda \gg M$. Although ``Majoron" interactions of the form $\phi NN$ are allowed by all the symmetries of the setup, the coefficients of such operators must be tuned to ensure that dangerous Majorana masses $\sim r_i NN$ do not arise to washout the asymmetry. 
Here the $y_\nu HLN$ term is merely an interaction which allows $N$ to decay above the weak scale; not as a source of all active neutrino masses.\footnote{After EWSB, for each generation one linear combination of $\nu_L$ and $N^c$ spinors acquires a Dirac mass with $N$, while the other remains massless. Thus, this simple model does not fully account for active neutrino masses, which are assumed to arise from additional interactions.}  

Assuming $N$ and $N^c$ are produced thermally during reheating, their dynamics give rise to an additional symmetric population of $\phi$ which is distinct
from the cold pre-inflationary condensate.  So long as the $\phi N \leftrightarrow \phi N$ scattering rate exceeds the Hubble rate, the symmetric 
$\phi$ population will thermalize with the radiation bath. However, this leading order process only exchanges energy, but not $B-L$ number. To communicate the asymmetry in the $\phi$ condensate, we demand that 
the rate  for the leading $B-L$ exchanging process $\phi N \to \phi \phi^* N^c$ (depicted in Fig. 1) exceeds Hubble before the $N$ decay to 
SM states. In the $ \Lambda \gg T \gg  M$ limit, the cross section for this interaction is approximately
$\sigma \sim T^{4}/ 8 (4\pi)^3  \Lambda^6$
%, where we have approximated the 3 body phase space as $T^2/4(4\pi)^3$. 
so the typical $\phi$ from the nonthermal pre-inflationary population
will exchange $B-L$ number with an $N$ in the thermal bath so long as 
$n_N \sigma \gtrsim H$, which requires
\be\label{eq:thermal-criterion}
 \Lambda \lesssim  0.14 \left( \frac{m_{Pl} T_R^5}{\sqrt{g_*}} \right)^{1/6}  \! \! \!   =  ~ 1.2 \times 10^3 \, \TeV \, \left( \frac{ T_R }{50\, \TeV}  \right)^{5/6}\!\!\! ,~~~~~~~
\ee
 assuming $g_* = 100$.  
Satisfying the condition in Eq.~(\ref{eq:thermal-criterion}) also guarantees thermalization between the $N$, $N^c$, and
the full $\phi$ population, including both symmetric-thermal and asymmetric condensate components.\footnote{For a more detailed discussion of (pseudo)scalar condensate thermalization in the context
of axions, see \cite{Masso:2002np,Braaten:1991dd,Bolz:2000fu} which perform more careful calculations. Although a similarly detailed 
treatment is necessary to determine the exact thermalization requirement here, 
Eq.~(\ref{eq:thermal-criterion}) suffices as a rough estimate to show that this can easily take place. 
Note that, unlike axions produced via misalignment, here the asymmetric $\phi$ population is always present.
 For a discussion of this issue see \cite{Masso:2002np}, which addresses axion thermalization {\it except} in the case of misalignment production, for which 
the population is not present before the QCD phase transition. }   After the asymmetry is transferred, $N\to h\nu_L$ decays transmit lepton number to the active neutrinos.

%%%%%%%%%%%%%%%%%%%%%%%%%%%%%%%%%%%%%%%%%%
%						Transfer scattering operator
%%%%%%%%%%%%%%%%%%%%%%%%%%%%%%%%%%%%%%%%%%

\begin{figure}[t!] 
\includegraphics[height=0.2\textwidth]{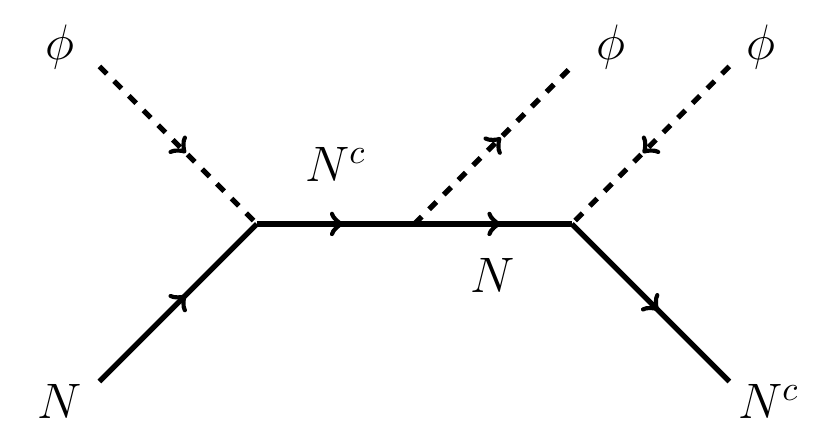}
\caption{Scattering process that exchanges lepton number between the pre-inflationary $\phi$ population
to the $N$ $N^c$ system after reheating when the condensate evaporates away. Lower order interactions (e.g. $\phi N \to \phi N$) can exchange
energy, but do not transfer $B-L$ number.}
\label{fig:betadecay} 
\end{figure}

%%%%%%%%%%%%%%%%%%%%%%%%%%%%%%%%%%%%%%%%%%
%%%%%%%%%%%%%%%%%%%%%%%%%%%%%%%%%%%%%%%%%%

For this particular example, we demand that this transfer occur before the electroweak phase transition so that sphalerons can distribute the asymmetry to the quarks; however this requirement is relaxed if $\phi$ also carries baryon number.  Following the procedure in \cite{Harvey:1990qw,Weinberg:2008zzc}, assuming
  the interaction in Eq.~(\ref{eq:transfer-lepton}) achieves thermal equilibrium before the electroweak phase transition, the baryon asymmetry is
\be\label{eq:YB}
Y_{ B}  =  \frac{28 \,  Q_\phi Y_{ \phi} }{22 \, Q_\phi^2 + 79} ~,~
\ee
where $Q_\phi$  is $\phi$'s $B-L$ number. It is clear that the asymmetry in $\phi$ from Eq.~(\ref{eq:Yphi}) is sufficient to generate a large baryon asymmetry. 

%%%%%%%%%%%%%%%%%%%%%%%%%%%%%%%%%%%%%%%%%%
%						Constraints
%%%%%%%%%%%%%%%%%%%%%%%%%%%%%%%%%%%%%%%%%%

\section{Constraints}

\noindent {\it Relativistic Degrees of Freedom}: A light $\phi$ particle thermalized with the SM can contribute to the effective relativistic degrees of freedom $N_{\rm eff}$ 
in the early universe. This bound can be evaded if the interaction Eq.~(\ref{eq:transfer}), which thermalizes the $\phi$ and SM sectors, decouples well before the CMB is formed, so
the $\phi$ temperature is  lower than that of the SM radiation at last scattering
 (for a review see \cite{Brust:2013xpv}). Scattering via $\phi \nu \leftrightarrow Z \nu$ keeps the $\phi$ population
 in thermal equilibrium until the $\phi$ decoupling temperature 
 \be 
T_{\phi,d}  \sim \!  2 \!     \times \! 10^3 \, {\rm TeV} \!   \left( \frac{10^{-5}}{y_\nu} \right)^{\! 4/3}  \!\!    \left( \frac{\Lambda}{10^3 \, \TeV} \right)^{\! 2/3}\!\!  \left( \frac{M}{10^2 \, \GeV} \right)^{\!2} \!\!   ,~~
 \ee
 which is sufficiently high that $g_{*}(T_{\phi,d}) \sim 100$. At the neutrino decoupling temperature $T_{\nu,d} \sim $ MeV, $g_{*}(T_{\nu}) \sim 10$,
 so at subsequent times
 \be
\frac{ T_{\phi} }{  T_{\nu}  }  = \left(   \frac{ g_{*}(T_{\phi,d })  }{   g_{*}(T_{\nu,d })  } \right)^{1/3} \! \sim 0.5  \,   \to  \,  
\delta N_{\rm eff} \sim \frac{  \rho_{\phi}   }{   \rho_{\nu}  } \sim 0.06,~~~~~~
 \ee
which is easily compatible with the Planck result $N_{\rm eff} = 3.30\pm0.27$ \cite{Ade:2013zuv}.

%%%%%%%%%%%%%%%%%%%%%%%%%%%%%%%%%%%%%%%%%%
%						Big operator diagram
%%%%%%%%%%%%%%%%%%%%%%%%%%%%%%%%%%%%%%%%%%

\begin{figure}[t!]
\hspace{-0.5cm}\includegraphics[height=0.3\textwidth]{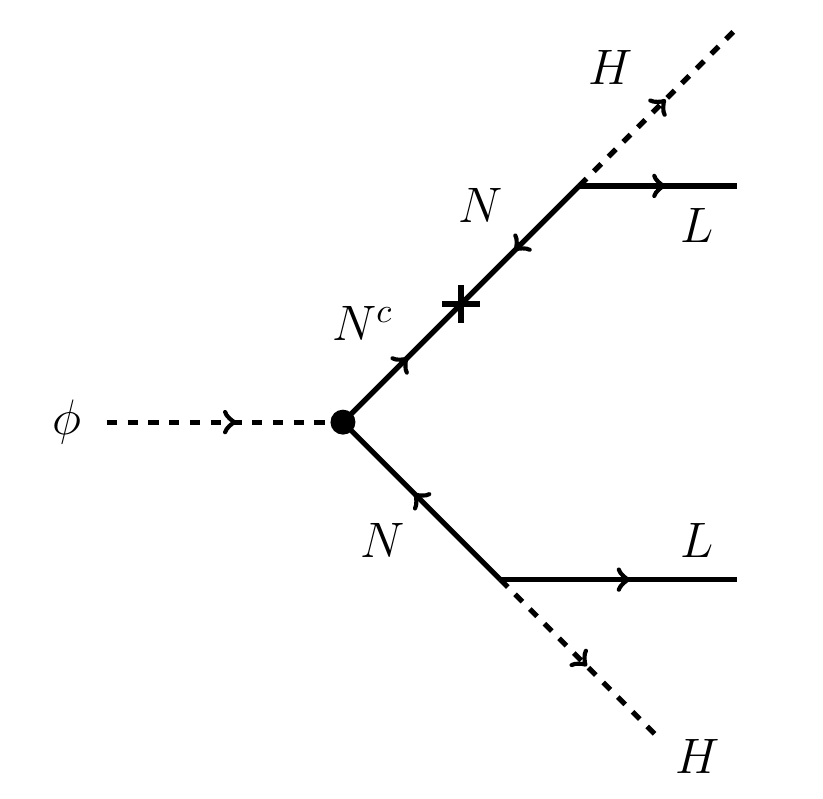}
\caption{Tree level interaction between $\phi$ and colorless doublets for the example model in Eq.~(\ref{eq:transfer}). Integrating out $N$ and $N^c$ gives 
rise to the operator in Eq.~(\ref{eq:bigoperator}).    }
\label{fig:bigdiagram}
\end{figure}

%%%%%%%%%%%%%%%%%%%%%%%%%%%%%%%%%%%%%%%%%%
%%%%%%%%%%%%%%%%%%%%%%%%%%%%%%%%%%%%%%%%%%

\medskip 
 
\begin{figure}[t!]
\hspace{-0.5cm}\includegraphics[height=0.2\textwidth]{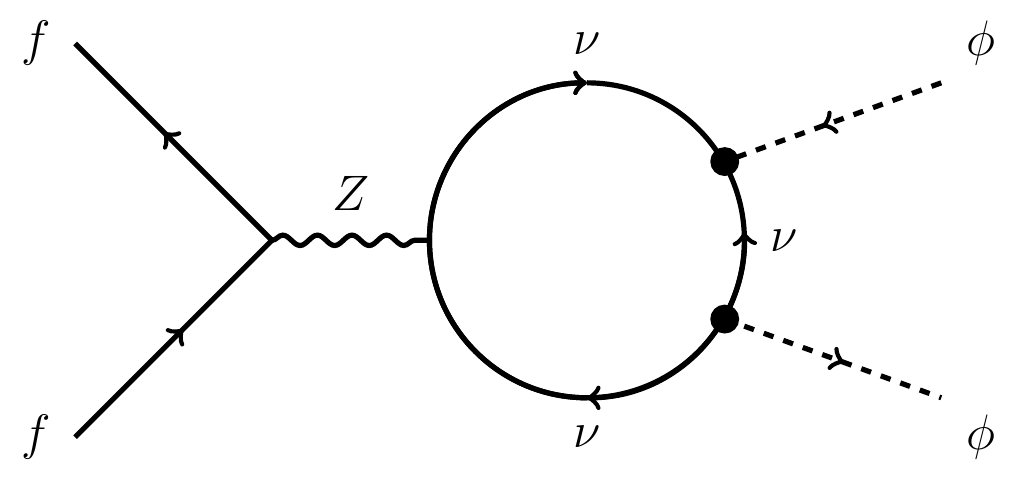}
\caption{Representative diagram contributing to apparent violation of the equivalence principle due to double $\phi$ exchange from the example interaction in Eq.~(\ref{eq:transfer}) after $N$ and $N^c$ have been integrated out and electroweak symmetry has been broken. Note
that this theory perturbatively preserves lepton number, so single-$\phi$ exchange diagrams are not generated in scattering processes with 
SM fermions $f$.}
\label{fig:longrange}
\end{figure}

\noindent {\it Equivalence Principle}: The first operator in  Eq.~(\ref{eq:transfer}) couples a nearly massless $\phi$ field to some 
combination of SM charged fields $\hat{\pazocal O}^{\mu}_{B-L}$, which can induce apparent violation of the equivalence principle through loop level interactions
with SM particles. 
For energies below the scale $M$, integrating out the $N$ and $N^c$ 
induces the effective interaction 
\be\label{eq:bigoperator}
{\cal L}_{\rm eff}  \supset   \frac{  c_{ij}  {y^*_{\nu}}^{i k }{y^*_{\nu}}^{j \ell } }{\Lambda M^3} \, (\partial_\mu \phi ) \, H^\dagger L_k^\dagger \bar \sigma^\mu \sigma^\nu \partial_\nu (H^\dagger L_\ell^\dagger) + h.c.~,~~~
\ee
from the diagram depicted in Fig.~\ref{fig:bigdiagram}, where $c_{ij}$ is an order one number introduced to restore the flavor 
dependence suppressed in Eq.~(\ref{eq:transfer}).
After electroweak symmetry breaking (EWSB), this interaction becomes  
\be\label{eq:EOMoperator4}
{\cal L}_{\rm eff} \supset \frac{  c_{ij}  {y^*_{\nu}}^{i k }{y^*_{\nu}}^{j \ell } v^2 }{\Lambda M^3}  (\partial_\mu \phi) \, {\nu}^\dagger_k \, \bar \sigma^{\mu} \sigma^\nu \partial_\nu \nu^\dagger_\ell  + h.c.~,~~~~~~~~
\ee
which gives rise to $\phi$ interactions with SM quarks and leptons at loop level through the diagram shown in Fig.~\ref{fig:longrange} (and
others of similar order). The leading non relativistic behavior for this diagram scales as $G_F |\vec q\,|^2$, where $q$ is the momentum transfer, which vanishes in the static limit. However, for $| \vec{ q} \,  |\ll m_\nu\sim 0.1\, \eV$, integrating out the neutrinos and connecting the scalars to a fermion current as shown in Fig. \ref{fig:scalarloop} gives rise to a  static force between SM fermions at 3 loop order. Working in the effective theory below $m_\nu$, the coefficient of the leading contribution to a
 Yukawa potential raising from this interaction scales as $ \sim  (y_\nu v)^8 G_F^2 m_\nu^{12}/ (1024 \pi^7 M^{12} \Lambda^4)$, which contributes negligibly to equivalence principle tests \cite{Schlamminger:2007ht}; see
the appendix for a discussion.

\medskip

\noindent{\it Supernova Cooling}: In analogy with axion like particles, efficient $\phi$ production in a supernova can potentially 
accelerate cooling during core collapse \cite{Turner:1987by}. However, $\phi$ only couples to neutrinos at tree level,
so even ignoring the significant prefactor suppression from Eq.~(\ref{eq:bigoperator}), radiative $\phi$ emission from neutrino nucleon scattering (e.g. $\nu n \to \nu n \phi$) does not accelerate the overall cooling rate since neutrinos are responsible for the conventional cooling mechanism (in the neutrino free streaming limit, this process is completely shut off). Scalar pair production off charged fermions through the loop process depicted 
in Fig. 3 is more suppressed relative to radiative electroweak neutrino production, so this contributes negligibly to the overall energy loss.

%%%%%%%%%%%%%%%%%%%%%%%%%%%%%%%%%%%%%%%%%%
%						Tuning
%%%%%%%%%%%%%%%%%%%%%%%%%%%%%%%%%%%%%%%%%%

\section{Tuning Considerations}
Since the $\phi$ radial displacement is super-Planckian and remains comparably large
well after inflation, this value must be sequestered from SM fields.  For any scalar field, there is an irreducible tuning in the Higgs portal coupling $  H^\dagger H |\phi |^2$ which must be chosen to prevent  destabilizing  $r_i^2 H^\dagger H$ corrections; indeed, this is an extreme realization of the usual electroweak hierarchy problem. Since no symmetry forbids this operator, its coefficient 
can only be naturally small if $H$ and $\phi$ are localized away from each other in a higher dimensional model. In such a scenario, the Higgs portal coefficient can be exponentially suppressed by wave function overlap, but additional model building may be required for $\phi$ to thermalize with the SM and transfer its asymmetry after reheating. At the nonrenormalizable level, there is no symmetry forbidding $B-L$ violating ``Planck slop" with potentially large insertions of $r_i$; if these arise
in a full theory of quantum gravity, they must be similarly tuned to prevent washout after reheating. 

The scalar $\phi$ must also be extremely light to decouple its number and energy densities, so its
 mass and interactions must also be exponentially tuned in this setup. For the particular example in Eq.~(\ref{eq:transfer}), 
 the coefficients of $B-L$ preserving ``Majoron" interactions $\phi NN$ and $\phi^* N^cN^c$ must also be highly tuned to ensure that $r_i$ does not give rise 
 to a large Majorana masses; such terms introduces dangerous $B-L$ breaking washout processes that can erase the asymmetry
 after reheating. 
    
If, instead, $\phi$ is interpreted as an NGB, there is partial protection from some of the tuning
presented in the example given above. Most favorably, its couplings  automatically involve derivatives, so there is no 
tuning required to protect SM operators from destabilizing $r_i$ insertions. Furthermore, the model in \cite{Kaplan:2015fuy} 
demonstrates that, for an ensemble of $p$ symmetry breaking scalar fields with suitable quartic interactions, the compact
field range of an NGB can be exponentially enhanced by factors as large as $\sim 3^p$ relative to the usual maximum excursion $\sim 2\pi f$, where $f$ is the symmetry breaking scale. For $f \sim m_{Pl}$ and $p=100$,  this enhancement allows an NGB to have a field range as large as $\sim 10^{67}$ GeV, which can generate an asymmetry sufficiently large to survive 60 e-folds of dilution and yield the observed asymmetry after thermalization with the SM-- see Eq.~(\ref{eq:Yphi}).

%%%%%%%%%%%%%%%%%%%%%%%%%%%%%%%%%%%%%%%%%%
%			             Long range precursor  -  neutrino loop
%%%%%%%%%%%%%%%%%%%%%%%%%%%%%%%%%%%%%%%%%%

 %%%%%%%%%%%%%%%%%%%%%%%%%%%%%%%%%%%%%%%%%%
%%%%%%%%%%%%%%%%%%%%%%%%%%%%%%%%%%%%%%%%%%

However, the NGB realization does not automatically  eliminate the Higgs portal tuning encountered in the fundamental scalar example above. Since exponentially enlarging the compact NGB field range requires a large population of natural, Planckian scalars, each of these $\sim 100$ fields must be prohibited from coupling to the $H^\dagger H$ bilinear and destabilizing the electroweak scale with Planck scale mass parameters. However, as discussed above, higher dimensional locality can exponentially reduce the Higgs portal coupling, so combined with the NGB realization, which protects the mass and ensures derivative interactions, there is a potential roadmap for a more natural theory.

%%%%%%%%%%%%%%%%%%%%%%%%%%%%%%%%%%%%%%%%%%
%						Discussion
%%%%%%%%%%%%%%%%%%%%%%%%%%%%%%%%%%%%%%%%%%

\section{Discussion}
In this letter we have presented a simple model to demonstrate that the baryon asymmetry can, in fact, arise from initial conditions in a universe that undergoes inflation. This 
is accomplished using a  light, bosonic
field with a large number density prior to inflation. At face value, the benchmark model requires severe  tuning to prevent super-Planckian field values from destabilizing the electroweak sector; however, we regard this setup merely as a proof of principle rather than as a compelling model of nature. If, contrary to the example model considered here, the asymmetry carrying field is an NGB in an extra-dimensional setup, it may be possible to build a natural theory that preserves many of the features discussed in this paper. However, doing so requires additional model building to 
engineer a super-Planckian field range for the asymmetry carrying field.

To be concrete in our discussion, we  have chosen a specific operator ansatz to communicate the $\phi$ asymmetry to the SM through the right-handed neutrino portal, but the key features of our discussion ff
do not depend on this choice. For instance, instead of using $\hat{\pazocal O}^{\mu}_{B-L} = {N^c}^\dagger \bar \sigma^\mu N$ in Eq.~(\ref{eq:transfer}), we could have assigned $\phi$ a net baryon number and transferred $B-L$ to quarks using singlet squarks from the Minimal Supersymmetric Standard Model via $\hat {\pazocal O}^{\mu}_{B-L} = (\partial^\mu \tilde u^c) \tilde d^c \tilde d^c$, with qualitatively similar results, up to different constraints introduced by this operator.  However, regardless of the particular choice of $\hat {\pazocal O}^{\mu}_{B-L}$, it is essential that the combination $\partial_\mu \phi \, \hat {\pazocal O}_{B-L}$ preserve 
$B-L$ number and that this symmetry is not broken by the interactions that resolve this operator. Indeed, aside from the initial  $\phi$ displacement before inflation, which spontaneously breaks $B-L$, all interactions in the example model from Eq.~(\ref{eq:transfer}) are $B-L$ singlets.

We note in passing that the effective operator in Eq.~(\ref{eq:EOMoperator4}) induces a new double $\beta$-decay signature, 
which contributes a {\it background} to neutrinoless double $\beta$ decay searches \cite{Osipowicz:2001sq}.
The final state for this $2n \to  2p 2e \phi $ process has different kinematics relative to conventional double $\beta$ decay  and 
may be distinguishable using electron momentum distributions. However,
since the model in Eq.~(\ref{eq:transfer}) does not fully account for neutrino masses, the overall rate depends on neutrino yukawa couplings, which
are not fixed in the setup described here. 

Finally, although this model does not address the electroweak hierarchy problem, it contains 
many of the ingredients found in the relaxion solution \cite{Graham:2015cka} ({\it e.g.} super Planckian field values for a light, slowly evolving field during inflation) so it would be interesting to see if the setup presented here can accommodate the cosmological relaxation of the electroweak scale as well.

%%%%%%%%%%%%%%%%%%%%%%%%%%%%%%%%%%%%%%%%%%
%				  	 	Scalar Loop
%%%%%%%%%%%%%%%%%%%%%%%%%%%%%%%%%%%%%%%%%%

\begin{figure}[t!] 
\includegraphics[height=0.2\textwidth]{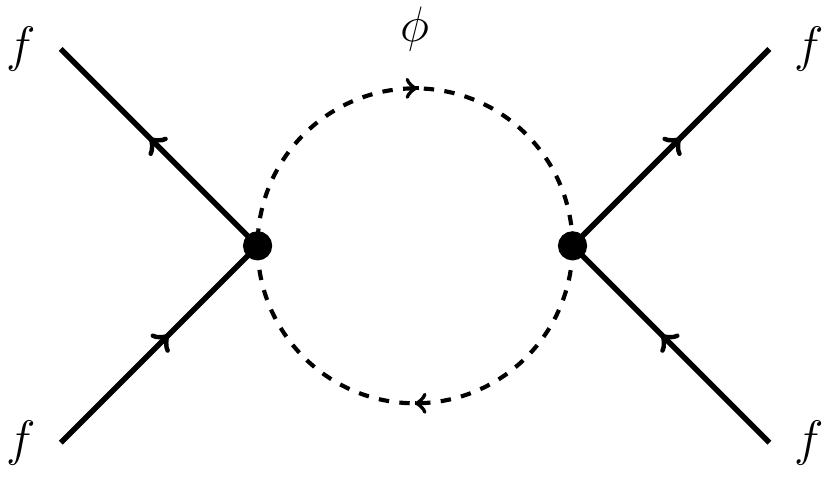}
\caption{Diagram contributing to a static long range force between SM fermions $f$ after integrating out the loop of neutrinos in Fig.~\ref{fig:longrange}.}
\label{fig:scalarloop} 
\end{figure}

%%%%%%%%%%%%%%%%%%%%%%%%%%%%%%%%%%%%%%%%%%
%%%%%%%%%%%%%%%%%%%%%%%%%%%%%%%%%%%%%%%%%%

\begin{acknowledgments}
{\it Acknowledgments} :  
We are  grateful to Cliff Burgess, Bogdan Dobrescu, Scott Dodelson, Paddy Fox, Roni Harnik, Anson Hook, Dan Hooper, Kiel Howe, Seyda Ipek,  David E. Kaplan, Jack Kearney, Zhen Liu, Jeremy Mardon, Stephen Martin, David Morrissey, Maxim Pospelov, Brian Shuve, and Flip Tanedo for many helpful conversations. Fermilab is operated by Fermi Research Alliance, LLC, under Contract No. DE-AC02-07CH11359 with the US Department of Energy.  
\end{acknowledgments}
 \medskip

\bigskip
\bibliography{PoissLep}

\vspace{1cm}

\section*{Appendix: Long Range Force Bounds}

The long range potential arising from the double $\phi$ exchange between SM fermions can be evaluated by taking the $q\to 0$ limit of the amplitude 
in Fig.~\ref{fig:scalarloop}. Although the sub diagram for this process shown in  Fig.~\ref{fig:longrange} is a momentum dependent interaction which
vanishes in the static limit, in the loop diagram these vertices can, instead, be sensitive to loop momenta which need not vanish
in this limit.  Although extracting the full functional dependence of this potential 
is beyond the scope of this paper, we can conservatively bound the leading contribution by evaluating 
\be \label{eq:potential}
\hspace{-0.1cm}V(\vec x) \sim  \hspace{-0.05cm} \frac{(y_\nu v)^8 G_F^2 m_\nu^6}{ (16 \pi^2)^2  \Lambda^4 M^{12}} \!  \int \!\! \frac{d^3 \vec q}{(2\pi)^3}  \int  \!\! \frac{d^4\ell}{(2\pi)^4} \frac{\ell^4 \, e^{i \vec q\cdot \vec x }  }{(\ell+\frac{q}{2})^2 (\ell-\frac{q}{2})^2} 
,~~~
\ee
where we have dropped the $q$ dependence from the numerator and extracted the leading $\ell^4$ piece while ignoring the spin structure, which can only 
suppress the long range behavior.
 The $\ell$ integral can be written in terms of a Feynman parameter and Euclidean momenta as
\be
\frac{i}{8\pi^2} \int_0^1 dx \int^{m_\nu}_0  \!\!\!  {\frac{d\ell_E  \, \ell_E^7}{   ( \ell_E^2   + \Delta)^2    }}~~~,~~~\Delta \equiv \frac{q^2 x}{2}  (1+ 2x)~,~~
\ee
where we have imposed a cutoff at $m_\nu$, the scale at which this effective theory breaks down. Evaluating the 
$\ell_E$ integral yields
\be \label{eq:rand}
\int_0^{m_\nu} \!\!\! {\frac{d\ell_E  \, \ell_E^7}{   ( \ell_E^2   + \Delta)^2    }} = \frac{1}{4} \biggl[ 3 m_\nu^4 + 6 \Delta^2 \log\left(\frac{\Delta + m_\nu^2}{ m_\nu^2}\right) \nonumber \\ \hspace{3cm}- 6 \Delta m_\nu^2  - \frac{2 m_\nu^6}{\Delta + m_\nu^2}\biggr]~,~~~
\ee
where only the last piece gives nontrivial dynamics in the static $q \to 0$ limit; all other terms are either zero or constant (the latter equivalent to a contact interaction in position space).
Performing the Feynman integral on the last term in Eq.~(\ref{eq:rand}) gives 
\be
-\frac{m_\nu^6}{2} \int_0^1 \frac{dx  }{  \frac{q^2 x}{2}(1+2x) + m_\nu^2    } \simeq \frac{8 m_\nu^6}{ q^2 - 16 \,m_\nu^2} ~,~~
\ee
so the non relativistic potential contains a dominant long distance contribution from
\be \label{eq:potential2}
V(\vec x) &\sim&  \hspace{-0.05cm} - \frac{  (y_\nu v)^8 G_F^2 m_\nu^{12}}{ 256\, \pi^6  \Lambda^4 M^{12}} \!  \int \!\! \frac{d^3 \vec q}{(2\pi)^3} 
\frac{e^{i \vec q\cdot \vec x}  }{ |\vec q\,|^2 + 16 \, m_\nu^2} \nonumber \\ 
\nonumber \\ 
&=&~~ - \frac{1}{4 \pi} \, \frac{ (y_\nu v)^8 G_F^2 m_\nu^{12}}{ 256 \, \pi^6  \Lambda^4 M^{12}} \, \frac{  e^{- 4 m_\nu r} }{r}~~,
\ee
which is negligibly small compared to Newtonian gravity in our regime of interest $\Lambda, M \gg v$, so fifth force searches  \cite{Schlamminger:2007ht}  do not constrain this scenario.

\end{document}